\documentclass[aps,pra,twocolumn,superscriptaddress,showpacs,floatfix]{revtex4-1}
\usepackage[enableskew]{youngtab}
\usepackage{graphicx,ytableau}
\usepackage{epstopdf,color}
\usepackage{amsmath,amssymb,xspace,epsfig,float,multirow}
\usepackage[utf8]{inputenc}
\begin{document}
\title{Finite temperature contact for a SU(2) Fermi gas trapped in a 1D harmonic confinement}
\date{\today}
\author{P. Capuzzi}
\email{capuzzi@df.uba.ar}
\affiliation{Departamento de Fisica, Universidad de Buenos Aires, Argentina}
\author{P. Vignolo}
\email{Patrizia.Vignolo@inphyni.cnrs.fr}
\affiliation{Universit\'e C\^ote d'Azur, CNRS, Institut de Physique de Nice,
1361 route des Lucioles
06560 Valbonne, France}
\begin{abstract}
  We calculate the finite-temperature Tan's contact for $N$ SU(2)
  fermions, characterized by repulsive contact interaction,
  trapped in a 1D harmonic confinement within a local density
  approximation on top of a thermodynamic Bethe Ansatz. The Tan's
  contact for such a system, as in the homogeneous case, displays a
  minimum at a very low temperature.  By means of an exact canonical
  ensemble calculation for two fermions, we provide an explicit
  formula for the contact at very low temperatures that reveals that
  the minimum is due to the mixing of states with different exchange
  symmetries.  In the unitary regime, this symmetry blending
  corresponds to a maximal entanglement entropy.

\end{abstract}
\maketitle
\section{Introduction}
The last two-decade progress in the manipulation and detection of
ultracold atoms has made this system one of the paradigms for quantum
simulators \cite{Blume2012,Sowinski2019}.
Indeed, it is possible to deal with bosons and/or
fermions, realise low dimensional systems
\cite{Moritz2003,Hadzibabic2006}, tune interactions by exploiting
Feshbach resonances \cite{Chin2010}, vary the number of spin
components \cite{Pagano2014}, and vary the number of particles from
many to few \cite{Wenz2013} down to the two-particle limit
\cite{Zurn2012}.  In particular, one-dimensional (1D) Fermi gases are
ideal quantum simulators for the exploration of quantum magnetism
\cite{Liao2010,Pagano2014,Deuretzbacher2014,murmann2015,He2016,Zinner2016}.
Recently, it has been
shown that the spin-resolved density profiles are not unambiguous
observables for the magnetic structure of $\kappa$-component 1D
SU($\kappa$) fermionic systems \cite{Decamp2016a}, while the Tan's
contact values for each of the components are \cite{Decamp2016}.
Namely, different symmetry configurations of a quantum many-body state
can correspond to the same spin-resolved density profiles, but there
is a one-to-one correspondence between each symmetry configuration and
its Tan's contact value \cite{Decamp2016}.  Tan's contact is an
observable that embeds the information about how particles can
approach each other taking into account the presence of all the other
particles in the system \cite{Tan2008a,Tan2008b,Tan2008c,Zwe11}.
Therefore, it depends on the number of particles, spin components,
interaction strength, temperature and on the external confinement.
Unlike the case of 1D homogeneous systems that can be exactly solved
\cite{LiebLin,Yang1967,Gaudin1967}, the 1D harmonically trapped
systems, that correspond to the usual experimental situation, cannot
be exactly solved for any interaction strength, temperature or number
of particles \cite{Sowinski2013,Gharashi2013,Volosniev2014}.
However, one can exploit energy scaling properties in
the thermodynamic limit to determine the contact for any (large)
number of particles by calculating it for a relative small number of
particles. This has been shown for repulsive bosons and
multi-component fermions at zero temperature
\cite{Olshanii03,MassignanPaarish,Gharashi2015,Lewenstein-Massignan,Matveeva2016,Decamp2016},
and for Lieb-Liniger bosons at finite temperature
\cite{Xu2015,Yao2018}. Moreover, for such systems, it has been shown
that the finite-interaction contact divided by the contact at the
unitary limit, for the same number of particles and temperature, is
(almost) a universal function even for a few particles.  This means
two things: first, that the $N$-dependency is almost completely
contained into the contact calculated at the unitary limit
\cite{Rizzi2018,Hebert2019}, which in turn can be exactly calculated
\cite{Vignolo2013,Yan2013,Yan2014,Hebert2019}; secondly, that a simple
two-body calculation at finite interactions and temperature is enough
to provide the contact for any number of particles with high accuracy
\cite{Rizzi2018,Hebert2019}.

The study of thermal repulsive multi-component fermions is much more
complex than a simple thermal Lieb-Liniger gas.  Indeed, the Bethe
Ansatz description for the homogeneous system provides an infinite
number of coupled equations \cite{Schlottmann93}.  At finite
temperature, the Lieb-Mattis theorem \cite{LiebMattisPR}, assuring
that the spatial wavefunction for the ground state is the most
symmetrical possible, does not hold any more. Different spin states
mix and the contact presents a minimum at low, finite temperature
that is more pronounced in the strong-interacting limit
\cite{Patu2016}.

In this paper we perform a finite-temperature local density
approximation (LDA) on the Bethe Ansatz solution for a SU(2) Fermi gas
\cite{Patu2016}, namely a two-component gas where each component has
the same mass and experiences the same external potential, and confirm
that the contact presents a well-defined minimum in the trapped gas
that is not washed out by inhomogeneity effects. This feature could be
observed not only in the momentum distribution tails, but also, for
instance, in the behaviours of the pair correlation function and the
loss rate in a mixture \cite{Sebastien2017} as a function of the
temperature.  Furthermore, due to the thermodynamic scaling being
independent of particle statistics, the LDA calculation for few
fermions provides the contact for any larger number of particles
\cite{Yao2018}. We compare this LDA result with a simple two-fermion
calculation. These two curves give upper and lower bounds for the
contact for any $N$, at corresponding rescaled interaction and
temperature \cite{Hebert2019}.  The two-fermion calculation also
allows us to enlighten the mechanism underlying the appearance of an
exchange symmetry mixing as a function of the temperature. We examine
the presence of this thermally driven symmetry blending in two
quantities connected to the one-body density matrix: the momentum
distribution and the von Neumann entanglement entropy.  Comparison
with the results for two Lieb-Liniger bosons and for the two
non-interacting fermions show that by increasing the temperature, the
two-boson momentum distribution hybridize with that of non-interacting
fermions.  At small interaction strength, we find an analogue behavior
for the von Neumann entanglement entropy: for two interacting fermions
such an entropy is in between that for two indistinguishable
interacting bosons and that for two non-interacting fermions.
However, at large interactions, the entanglement entropy for two
fermions grows very rapidly with temperature, exceeding the entropy of
the two non-interacting fermions.  This means that, in the strongly
interacting regime, the symmetry blending corresponds to a maximal
entanglement entropy, the symmetric and antisymmetric spin
configurations becoming energetically equivalent.

The manuscript is organized as follows.  The model for the trapped gas
is introduced in Sec. \ref{themod}, while its thermodynamical
description in the grand canonical ensemble is given in
Sec. \ref{thermo}.  The two-fermion calculation for the contact is
detailed in Sec. \ref{two-cont}. The momentum distribution and
entanglement entropy are discussed in Sec. \ref{sec:rho1}.  Finally,
Sec. \ref{concl} concludes the manuscript.


\section{The model: the harmonically trapped Yang-Gaudin gas}
\label{themod}
We consider a system of $N$ fermions of equal mass $m$, divided into
$2$ species with the same population.  We assume that the two
components are subjected to the same harmonic potential
$V (x) = m\omega^2x^2/2$, and that fermions belonging to different
species interact with each other via the contact potential
$v(x- x')=g\delta(x-x')$, where $g$ is the interaction strength, and
$\delta(x)$ is the Dirac delta function. The total Hamiltonian reads
\begin{equation}
  \mathcal{H}=\sum_{i=1}^N\left[-\dfrac{\hbar^2}{2 m}
    \dfrac{\partial^2}{\partial x_i^2}+\dfrac{1}{2}m\omega^2x_i^2\right]
  +g\sum_{i<j}\delta(x_i-x_j).
\end{equation}

This model is exactly solvable in the absence of harmonic confinement
\cite{Yang1967,Gaudin1967,Schlottmann93}, in the Tonks limit
$g\rightarrow\infty$ in presence of harmonic confinement
\cite{Decamp2016a,Decamp2016}, or for two particles for any
interaction.
\section{Tan's contact for $N$ SU(2) fermions}
\label{thermo}
\begin{figure}
  \begin{center}
    \includegraphics[width=1.\linewidth]{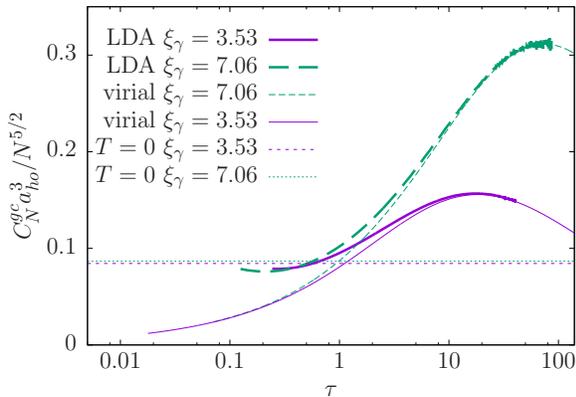}
  \end{center}
  \caption{\label{fig1}Rescaled grand-canonical contact
    $C_N^{gc}a_{ho}^3/N^{5/2}$ as a function of the reduced
    temperature $\tau$. Thick lines: LDA calculation, Eq.~(\ref{LDA}).
    Thin lines: virial expansion, Eq.~(\ref{virial}). Horizontal lines:
    zero-temperature values \cite{Decamp2016}. Solid violet curves:
    $\xi_\gamma=3.53$.  Dashed green curves:
    $\xi_\gamma=7.06$.}
\end{figure}

Thermodynamics of the 1D multicomponent Fermi gas with a
delta-function interaction is described by an infinite set of coupled
equations \cite{Schlottmann93}, that thus are numerically difficult to
implement.  However, for the case of a SU(2) gas, P\^a\c{t}u and
Kl\"umper \cite{Patu2016} have proposed an efficient thermodynamic
description that reduces the infinite set to two coupled integral
equations.  In such a frame, the thermodynamic grand-potential density
can be written
  \begin{equation}
    \Omega_h=-\dfrac{1}{2\pi\beta}\int {\rm d}k\,\left[\ln(1+e^{-\beta\epsilon_1(k)})+
      \ln(1+e^{-\beta\epsilon_2(k)})\right]
      \label{grandpot}
  \end{equation}
  where $\epsilon_1$ and $\epsilon_2$ satisfy the two following
  coupled integral equations over the wavevector $q$ (with
  $\alpha\rightarrow 0^+$):
  \begin{equation}
    \begin{split}
      \epsilon_1(k)&=\frac{\hbar^2k^2}{2 m}-\mu-\frac{c}{2\pi\beta}
      \int \frac{{\rm
          d}q\ln(1+e^{-\beta\epsilon_2(k)})}{(k-q-i\alpha)(k-q-i\alpha-ic)}
      \\
      \epsilon_2(k)&=\frac{\hbar^2k^2}{2 m}-\mu-\frac{c}{2\pi\beta}
      \int \dfrac{{\rm d}q 
        \ln(1+e^{-\beta\epsilon_1(k)})}{(k-q+i\alpha)(k-q+i\alpha-ic)}
    \end{split}
    \label{triste}
  \end{equation}
  with $\beta=1/k_BT$ and $c=mg/\hbar^2$.  P\^a\c{t}u and Kl\"umper
  have shown that the contact density for the homogeneous system
  \begin{equation}
 C_h=-\dfrac{m^2}{\pi\hbar^4}\dfrac{\partial\Omega_h}{\partial g^{-1}},
  \end{equation}
  exhibits a minimum at a temperature of the order of
  $T_{0,h}=T_F/\gamma$, $T_F=\pi^2\hbar^2n^2/(2m k_B)$ being the Fermi temperature,
  $\gamma=mg/(\hbar^2 n)$ and $n$ the density. For $T\gg T_{0,h}$ the
  spin degrees of freedom are ``disordered''\cite{Cheianov2015}, i.e.
  the different spin states mix together, whereas the density degrees
  of freedom are unaffected until the temperature becomes of the order
  of $T_F$. With the aim to verify if this minimum is not washed out
  by inhomogeneity in trapped systems, we perform a LDA for the
  calculation of the contact for the harmonically trapped system.  We
  replace in Eqs. (\ref{triste}) the chemical potential $\mu$ with the
  local value $\mu-m\omega^2x^2/2$, and obtain a local grand-potential
  density $\Omega_x$ that depends on the position.  This approximation
  is valid in the limit of large interactions $g/(\hbar\omega a_{ho})\gg 1$
  and large number of particles $N\gg1$.
  The value of $\mu$
  for the trapped system is thus obtained by imposing the
  thermodynamic constraint for an average number of fermions $N$,
  \begin{equation}
 N=-\int {\rm d}x\,\dfrac{\partial \Omega_x}{\partial \mu}.
  \end{equation}
  The contact $C_{N,LDA}^{gc}$, in the grand-canonical ensemble,
  evaluated using the LDA, for a trapped system of an average number
  of $N$ fermions, can thus be readily calculated as
  \begin{equation}
    C_{N,LDA}^{gc}=-\int {\rm d}x\,
    \dfrac{m^2}{\pi\hbar^4}\dfrac{\partial\Omega_x}{\partial g^{-1}}.
    \label{LDA}
  \end{equation}
  It can be easily shown that, as for the case of a bosonic system
  \cite{Yao2018}, the contact obeys a scaling law with $N$
  \begin{equation}
\dfrac{C_{N,LDA}^{gc}}{N^{5/2}}=f(\xi_\gamma,\xi_T)=\tilde f(\xi_\gamma,\tau)
\label{scaling}
  \end{equation}
  where $f$ is a universal function of the reduced interaction
  strength $\xi_\gamma=a_{ho}/(|a_{1D}\sqrt{N}|)$ and of the ratio
  between the one dimensional scattering length and the de Broglie
  wavelength $\xi_T=|a_{1D}|/\lambda_T$, $\tilde f$ is a universal
  function of $\xi_\gamma$ and the reduced temperature
  $\tau= k_BT/(N\hbar\omega)= 2\pi\xi_T^2\xi_\gamma^2$.  The 1D
  scattering length is defined by $a_{1D}=-2\hbar^2/(mg)$ and the de
  Broglie wavelength by $\lambda_T=\sqrt{2\pi\hbar^2/(m k_B T)}$. Lengths
  are measured in units of the harmonic oscillator
  $a_{ho}=\sqrt{\hbar/m\omega}$. Due to the LDA, the scaling law
  (\ref{scaling}) is expected to be valid in the limit of large $N$
  only.  In Fig. \ref{fig1} we show the results for interaction
  strengths $\xi_{\gamma}=3.53$ (solid violet curves) and $7.06$
  (dashed green curves).
  The LDA curves are compared with the contact $C_{N,vir}^{gc}$
  obtained by means
  of the virial expansion
  \begin{equation}
  C_{N,vir}^{gc}=\dfrac{N^{5/2}}{\pi a_{ho}^3}\dfrac{\xi_\gamma}{\xi_T}
  \left(\sqrt{2}-\dfrac{e^{1/2\pi\xi_T^2}}{\xi_T}\,{\rm erfc}
  (1/\sqrt{2\pi}\xi_T)\right),
  \label{virial}
  \end{equation}
  that has been obtained analogously to the bosonic case
  \cite{Yao2018}.  Eq. (\ref{virial}) is valid in the limit of large
  interactions ($\xi_\gamma>1$) and high temperature ($\tau\gg 1$).
  As for the bosonic case, this function has a maximum at
  $\xi_T=0.485$, namely at $\tau=1.48\,\xi_\gamma^2$.
  
\section{Understanding the contact: two-fermions calculation}
\label{two-cont}
As pointed out in \cite{Patu2016}, Eq. (\ref{grandpot}) is not
analytical at $T=0$ and thus it is not possible to get a Taylor
expansion at low temperatures.  However, it is possible to obtain an
explicit expression of the contact for two fermions in the canonical
ensemble from the Helmholtz free energy $F$.
In this ensemble the contact $C^c_N$ for $N$ particles reads 
\begin{equation}
  \begin{split}
    C^c_N&=-\dfrac{m^2}{\pi\hbar^4}\langle\dfrac{\partial F}{\partial g^{-1}}\rangle\\
    &=-\dfrac{m^2}{\pi\hbar^4}\dfrac{\sum_i e^{-\beta E_i}\dfrac{\partial E_i}{\partial g^{-1}}}{\sum_i e^{-\beta E_i}}.
\end{split}
\end{equation}
with $E_i=E_{cm,\ell}+E_{rel,j}$, where only the relative energy
$E_{rel,j}$ is a function of $g$, while the center-of-mass energy
$E_{cm,\ell}$ isn't. Aiming at clarifying the different contributions
to the energy in the two-fermion calculation, let us first review the
case of two trapped Lieb-Liniger bosons \cite{Hebert2019}.
 
\subsection{Two identical bosons.} 

\noindent
For the trapped system composed by two identical bosons interacting
through a Dirac delta potential, the spectrum of the relative energy
is analytically known \cite{Busch98} and can be written as:
\begin{equation}
E_{rel,i}=\hbar\omega\left(\frac{1}{2}+\nu(i)\right)
\label{eq:Erel}
\end{equation}
where $\nu(i)$, with $i\ge1$, satisfies the relation
\begin{equation}
\dfrac{\Gamma(-\nu(i)/2)}{\Gamma(-\nu(i)/2+1/2)}=f(\nu(i))=-\dfrac{2\sqrt{2}}{g}\hbar\omega a_{ho},
\end{equation}
where $\Gamma(x)$ is the Gamma function \cite{Gradshteyn}.  In the
unitary limit $g\rightarrow\infty$, $\nu_\infty(i)=2i-1$, where $i$ is
a positive integer labelling the levels. This corresponds to the
fermionized regime.

The derivative ${\partial E_{rel,i}}/{\partial g^{-1}}$ can be written
as a function of $f(\nu)$:
\begin{equation}
\dfrac{\partial E_{rel,i}}{\partial g^{-1}}=-2\sqrt{2}(\hbar\omega)^2a_{ho}\left(\dfrac{\partial f}{\partial\nu}\right)^{-1}.
 \end{equation}
Thus the canonical contact for two bosons $C_{2b}^c$ reads
\begin{equation}
  \begin{split}
    C_{2b}^c&=\dfrac{1}{\pi a_{ho}^3}2\sqrt{2}\dfrac{\sum_i e^{-\beta \hbar\omega\nu(i)}\left(\dfrac{\partial f}{\partial\nu}\right)_i^{-1}}{\sum_i e^{-\beta \hbar\omega\nu(i)}}\\
    &=\dfrac{\sum_i e^{-\beta \hbar\omega\nu(i)}C_i}{\sum_i e^{-\beta \hbar\omega\nu(i)}},
    \end{split}
\end{equation}
where
\begin{equation}
  C_i=\dfrac{1}{\pi a_{ho}^3}2\sqrt{2}\left(\dfrac{\partial f}{\partial\nu}\right)_i^{-1}
  \label{eqci}
\end{equation}
is the ``zero-temperature contact'' relative to the energy level $i$. 
\subsection{Two SU(2) fermions (or bosons).}

For the case of two fermions with two spin projections, we have to
consider that given that the total wavefunction must be antisymmetric
against particle exchange, their spatial part can be either symmetric
for the antisymmetric singlet spin state ($s=0$), or antisymmetric for
the symmetric triplet spin state ($s=1$).  The spatially symmetric
case is equivalent to the bosonic case, namely
$E_{rel,i}^s=E_{rel,i}=\hbar\omega(\nu(i)+1/2)$ and $C_{i}^s=C_i$,
where $C_i$ has been given in Eq. (\ref{eqci}).  The antisymmetric
case is energetically equivalent to the Tonks limit for bosons where
$E_{rel}^a=\hbar\omega(\nu_\infty(i)+1/2)$, but the contact terms
$C_{i}^a$ are vanishing.  Therefore, the canonical contact for two
fermions $C_{2f}^c$ reads
\begin{equation}
  \begin{split}
    C_{2f}^c&=\dfrac{\sum_i (e^{-\beta \hbar\omega\nu(i)}C_{i}^s+e^{-\beta \hbar\omega\nu_\infty(i)}C_{i}^a)}{\sum_i (e^{-\beta \hbar\omega\nu(i)}+e^{-\beta \hbar\omega\nu_\infty(i)})}\\
    &=\dfrac{\sum_i e^{-\beta \hbar\omega\nu(i)}C_{i}}{\sum_i (e^{-\beta \hbar\omega\nu(i)}+e^{-\beta \hbar\omega\nu_\infty(i)})}.
    \label{can-2F}
\end{split}
  \end{equation}
     At $T=0$ the fermionic contact coincides with that of two
  indistinguishable bosons since the ground-state is totally
  symmetric, while at high temperature the contact is equal to half
  of the bosonic one since the symmetric and antisymmetric components
  have almost the same weight. In between these two limits, the
  contact goes through a minimum as for the thermodynamic limit.
  
\begin{figure}
  \begin{center}
    \includegraphics[width=1.\linewidth]{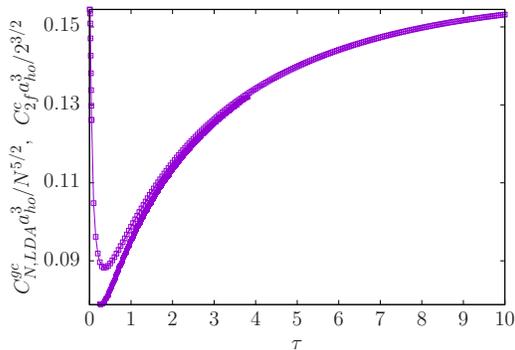}
     \end{center}
     \caption{LDA grand canonical contact $C_{N,LDA}^{gc}$ rescaled
       by $N^{5/2}$ (full symbols) and the canonical one $C_{2f}^c$
       (empty symbols) rescaled by $N^{3/2}(N-1)=2^{3/2}$ as functions
       of $\tau$, for the case
       $\xi_{\gamma}=3.53$.\label{fig-gc-c}}
\end{figure}

  In Fig. \ref{fig-gc-c}, we compare the LDA calculation
  with the exact two-fermion one for the case $\xi_{\gamma}=3.53$,
  by rescaling $C_{N,LDA}^{gc}$ by $N^{5/2}$ and
  $C_{2f}^c$ by $N^{3/2}(N-1)=2^{3/2}$, that are
  the high-temperature grand-canonical and canonical
  scaling factors for the contact \cite{Hebert2019}.
Indeed, the two curves collapse on the same curve at $\tau\gg1$.  On
the other hand, at low temperatures, these two scaling factors do not
hold for small number of particles \cite{Rizzi2018,Hebert2019} and the
two curves stay close but not superposed. However,
$C_{N,LDA}^{gc}/N^{5/2}$ and $C_{2f}^c/2^{3/2}$ provide lower and upper
bounds, respectively for both the rescaled grand-canonical contact
$C_N^{gc}/N^{5/2}$ for an average number $N$ of particles and the
rescaled canonical one $C_N^{c}/(N^{3/2}(N-1))$ for $N$ particles
\cite{Rizzi2018,Hebert2019}.

\subsubsection{$T\simeq 0$ behaviour}
\begin{figure}
\begin{center}
    \includegraphics[width=0.95\linewidth]{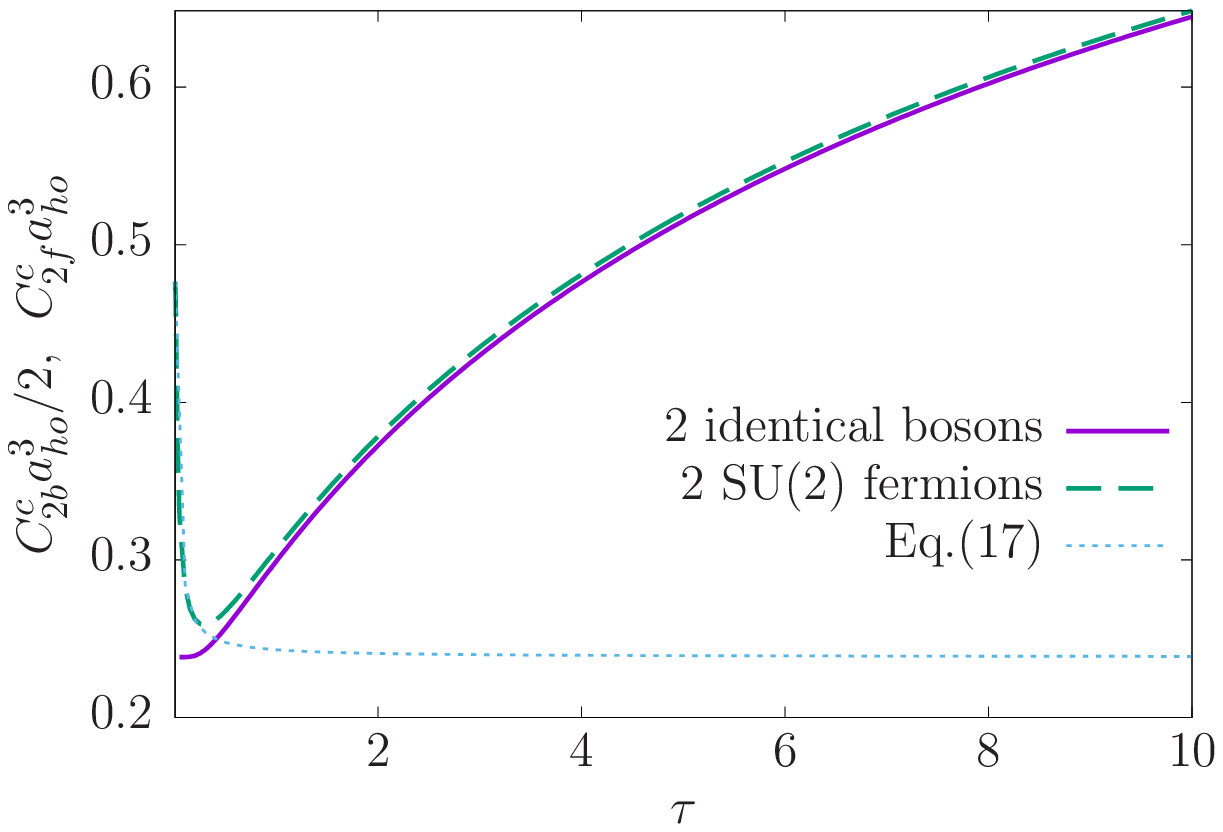}
  \includegraphics[width=0.95\linewidth]{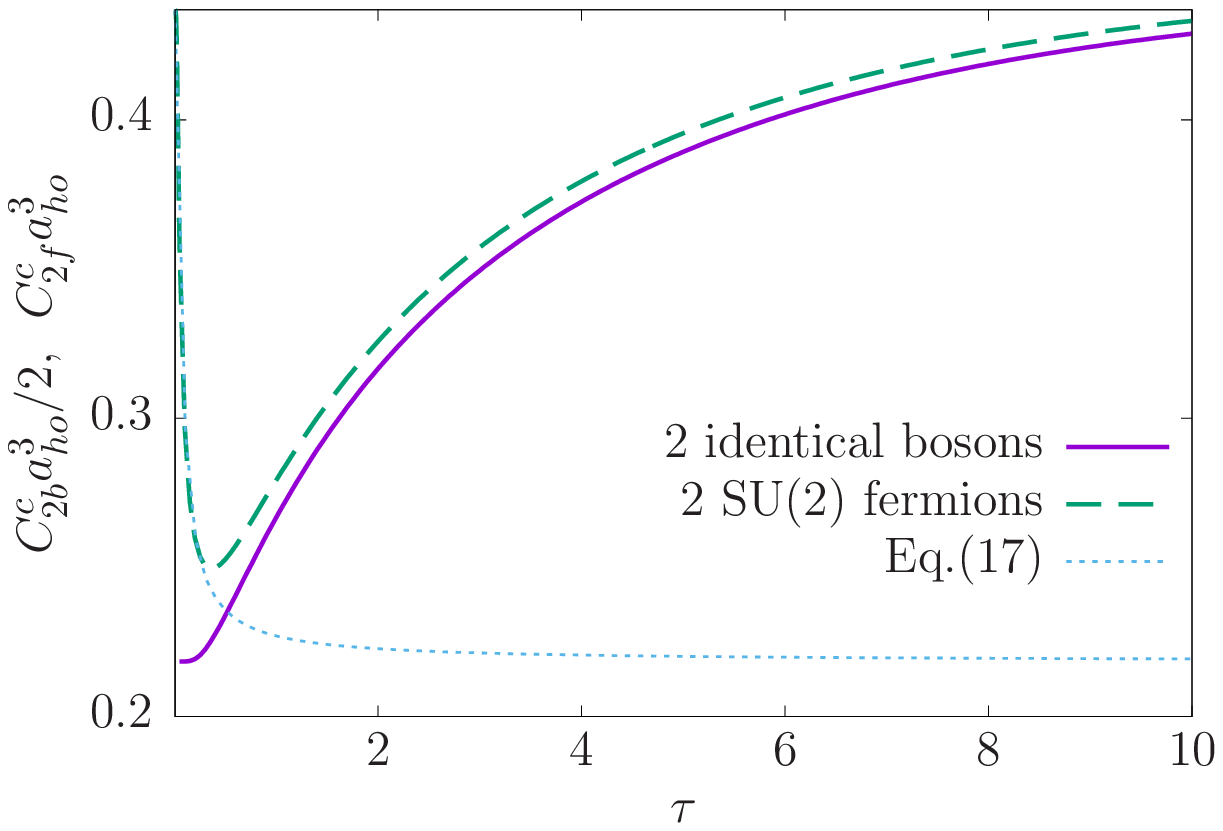}
  \caption{\label{fig1b}Two (identical) boson contact
    $C_{2b}^c/2$ (violet curve) and
    two SU(2) fermions contact $C_{2f}^c$ (green curve) as a function of $\tau$,
    for $g=20\hbar\omega a_{ho}$ (top figure)
    and $g=10\hbar\omega a_{ho}$ (bottom figure).
    The thin blue curve corresponds to Eq. (\ref{approx}).}
  \end{center}
\end{figure}

From Eq. (\ref{can-2F}), it is straightforward to show that, at $T=0$,
the two-fermions contact is equal to the two-identical-boson one. On
the other hand, at high temperature, the two-fermions contact is about
one-half of the bosonic one because the two terms in the denominator
are very close.  The high temperature regime is marked by $T\gg T_0$,
where $T_0=\hbar\omega[\nu_\infty(1)-\nu(1)]/k_B$ is the difference in
the ground-state energy between states with finite and infinite
interactions and is the analogue of $T_{0,h}$ for the trapped
system. Remark that (see \cite{Busch98})
$[\nu_\infty(i)-\nu(i)]\simeq [\nu_\infty(1)-\nu(1)]$, for any $i$.
In the limit of large interactions
\begin{equation}
  k_BT_0=\hbar\omega[\nu_\infty(1)-\nu(1)]\simeq
  -
  \dfrac{1}{g}\left.\dfrac{\partial E_{GS}}{\partial g^{-1}}\right|_{g\rightarrow\infty}
  \simeq\dfrac{\pi\hbar^4}{m^2}\dfrac{C_{1,\infty}}{g},
\end{equation}
where $E_{GS}$ is the zero temperature ground-state energy of the system,
and $C_{1,\infty}$ is the ground-state contact in the unitary limit.
In the same limit, one can find a simplified expression
for $C_{2f}^c$ at low temperatures
as follows
\begin{equation}
  \begin{split}
    C_{2f}^c(T\simeq 0)&\simeq \dfrac{e^{-\beta\hbar\omega\nu(1)}C_1}
    {e^{-\beta\hbar\omega\nu(1)}+e^{-\beta\hbar\omega\nu_\infty(1)}}\\
    &\simeq\dfrac{C_1}{1+e^{-\beta \pi \hbar^4 C_{1,\infty}/(g m^2)}}.\\
\end{split}
  \label{approx}
\end{equation}
Remark that $C_{2f}^c(T\simeq 0)$ is not an analytical function as
already pointed out in \cite{Patu2016}.  In Fig. \ref{fig1b} we show
the contact for two SU(2) fermions and one half of the contact of two
identical bosons for the cases $g=20\hbar\omega a_{ho}$ and
$g=10\hbar\omega a_{ho}$.  The minimum of the fermionic curves is
located at $T=T_{\min}\sim 5 T_0$ ($T_0/T_F=0.037$ for
$g=20\hbar\omega a_{ho}$, and $T_0/T_F=0.068$ for
$g=10\hbar\omega a_{ho}$).

\begin{figure}
\begin{center}
    \includegraphics[width=0.95\linewidth]{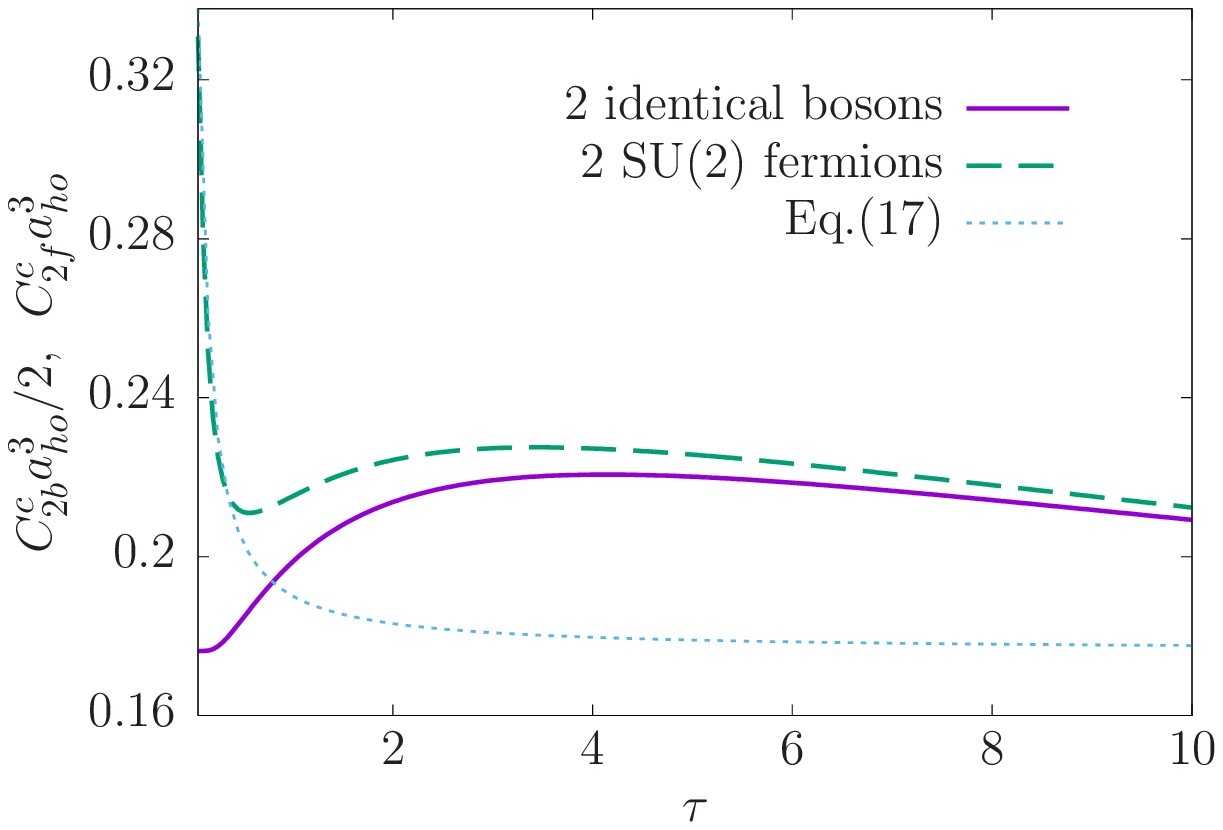}
  \includegraphics[width=0.95\linewidth]{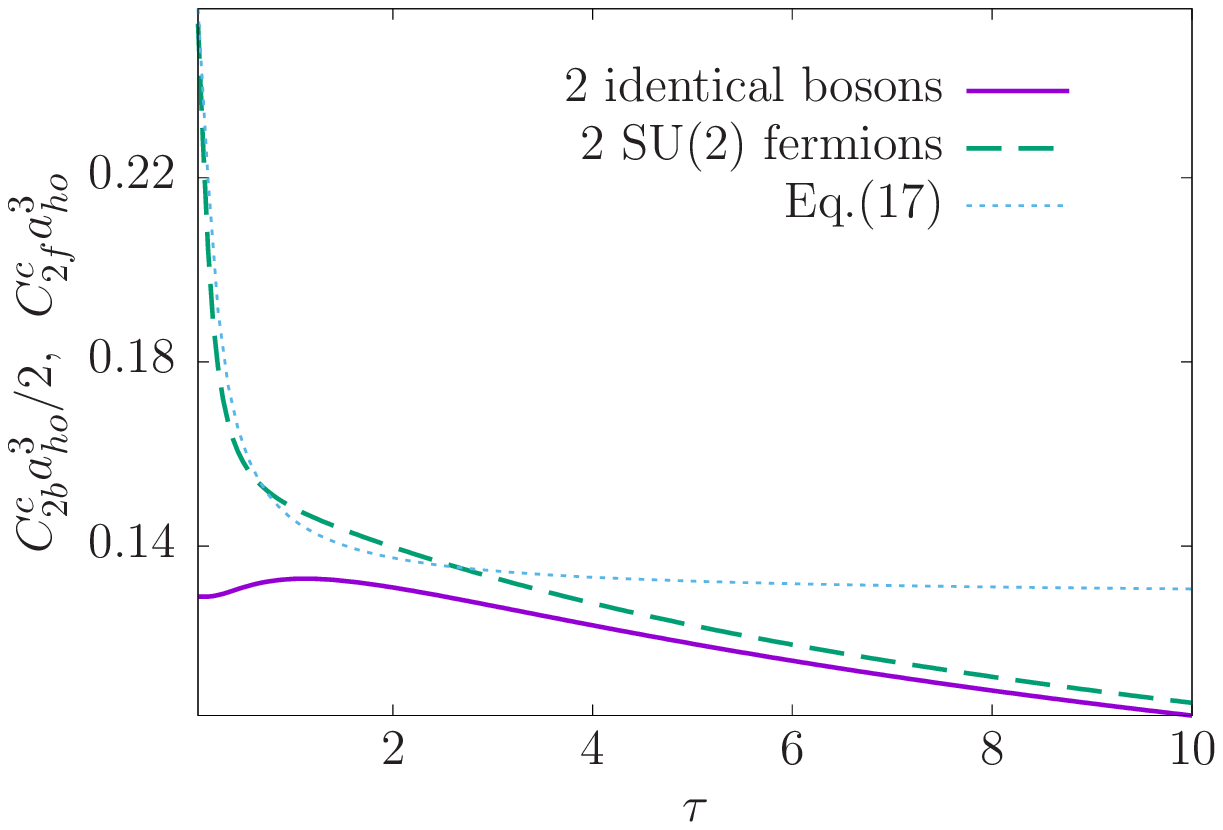}
  \caption{\label{fig2} Two (identical) boson
    contact $C_{2b}^c/2$ (violet curve) and
    two SU(2) fermions contact $C_{2f}^c$ (green curve) as a
    function of $\tau$, for $g=5\hbar\omega a_{ho}$ (top figure) and
    $g=3\hbar\omega a_{ho}$ (bottom figure).
    The thin blue curve corresponds to Eq. (\ref{approx}).}
  \end{center}
\end{figure}

In the strong interaction regime, $\xi_\gamma>1$ (large $g$), the
maximum of the contact is located at
$\tau=T_{\max}/T_F\simeq 1.48\xi_\gamma^2$ \cite{Yao2018}.  We can
expect that the minimum will disappear when $T_{\min}\simeq T_{\max}$,
which thus occurs at $g\simeq 3 \hbar\omega a_{ho}$.  In
Fig. \ref{fig2} we show the contact for two SU(2) fermions and two
identical bosons for the cases $g=5\hbar\omega a_{ho}$ and
$g=3\hbar\omega a_{ho}$.  At $g=5\hbar\omega a_{ho}$, the minimum and
the maximum are close, and they disappear at $g=3\hbar\omega a_{ho}$,
as expected.
The fact that the approximated expression (\ref{approx})
works quite well even at intermediate interactions is due to the
fact that $C_{1,\infty}/g$ is a good estimate of the difference
$\nu_\infty(1)-\nu(1)$ in such a regime too.

\subsubsection{Generalization of Eq. (\ref{approx})}

We now consider a system with $N$ SU(2) fermions.  Let $C_1$ be the
zero-temperature contact for the most symmetric state corresponding to
a Young tableau with a row of $N$ boxes
${\tiny \ytableausetup{mathmode,boxsize=1.5em,centertableaux}
  \setlength{\delimitershortfall}{-5pt}
  \begin{ytableau} 1&2&3&4&\dots&N 
  \end{ytableau}}$, and $\tilde C_1$ the zero-temperature contact for
the state corresponding to a Young tableau with one row with $(N-1)$
boxes and another row with one box
${\tiny \ytableausetup{mathmode,boxsize=1.5em,centertableaux}
  \setlength{\delimitershortfall}{-5pt}\begin{ytableau}
    1&2&3&\dots&N\!\!-\!\!1\\N \\\end{ytableau}}$
\cite{Hamermesh_book,Decamp2016}.
For such a system
\begin{equation}
  \begin{split}
C(T\simeq 0)&\simeq\dfrac{C_1+\tilde C_1 e^{-\beta\Delta E}}{1+e^{-\beta\Delta E}}\\
&\simeq\dfrac{C_1}{1+e^{-\beta\Delta E}},
\label{approx2}
  \end{split}
\end{equation}
where at the denominator we have neglected the contribution of
$\tilde C_1$ which is smaller than $C_1$ since it corresponds to a
less symmetric state.  The energy difference $\Delta E$, in the limit
of strong interactions, can be written as
\begin{equation}
\Delta E=\dfrac{ \pi\hbar^4}{m^2} \dfrac{C_{1,\infty}-\tilde C_{1,\infty}}{g}.
\end{equation}
The contact is proportional to the number of pairs that can interact:
$N(N-1)/2$ in $C_{1,\infty}$ and $(N-1)(N-2)/2$ in
$\tilde C_{1,\infty}$ (at least in the thermodynamic limit). Thus, one
finds that $C_{1,\infty}-\tilde C_{1,\infty}\simeq C_{1,\infty}2/N$.
Thus, for the case of $N$ fermions, Eq. (\ref{approx2}) takes the form
\begin{equation}
  \begin{split}
    C(T\simeq 0)&\simeq\dfrac{C_1}{1+e^{-\beta\Delta E}}\\
    &\simeq\dfrac{C_1}{1+e^{-\beta 2\pi\hbar^4 C_{1,\infty}/(gN m^2)}}\\
    &\simeq\dfrac{C_1}{1+e^{- \pi\mathcal{C}_{1,\infty}/(\tau\xi_\gamma)}}
  \end{split}
  \end{equation}
where $\mathcal{C}_{1,\infty}={C}_{1,\infty}/(N^{5/2}a_{ho}^3)$ is
the rescaled contact.
The usual thermodynamical scaling is recovered, since
in the thermodynamic limit $\mathcal{C}_{1,\infty}$ is a
universal function of $\tau$ \cite{Vignolo2013}.

\section{A thermally driven symmetry blending}
\label{sec:rho1}
The contact behaviour at low temperature is due to the exchange
symmetry mixing: at $T=0$ the only contribution to the contact
originates from the fully symmetric ground state, while with
increasing the temperature less symmetric states start to contribute
and the contact diminishes. The role of less symmetric states is
extremely clear for two fermions where the only possible states are
the fully symmetric and the fully antisymmetric with vanishing
contact.  Aiming at characterizing this symmetry blending process, we
have calculated the momentum distribution and the von Neumann
entanglement entropy for the two-fermion system. Both quantities can
be derived from the canonical one-body density matrix which in turn
can be written explicitly.
\subsection{The canonical one-body density matrix}
The canonical one-body density-matrix for two fermions reads
\begin{equation}
  \rho(x,x')=\dfrac{\sum_{i,j}e^{-\beta E_{i,j}^s}\rho_s^{i,j}(x,x')+
    \sum_{i<j}e^{-\beta E_{i,j}^a}\rho_a^{i,j}(x,x')}{\sum_{i,j}e^{-\beta E_{i,j}^s}+\sum_{i<j}e^{-\beta E_{i,j}^a}}
\label{eq:rdmT}
\end{equation}
where $E_{i,j}^s=E_{cm,i}+E_{rel,j}=\hbar\omega(i+\nu(j))$,
$E_{i,j}^a=\hbar\omega(i+j-1)$, with $i$ and $j$ $\ge1$.
$\rho_s^{i,j}(x,x')$ and $\rho_a^{i,j}(x,x')$ are respectively the
exchange symmetric and exchange antisymmetric contributions (see
Appendix).

\subsubsection{The momentum distribution.}

\begin{figure}
\begin{center}
    \includegraphics[width=1\linewidth]{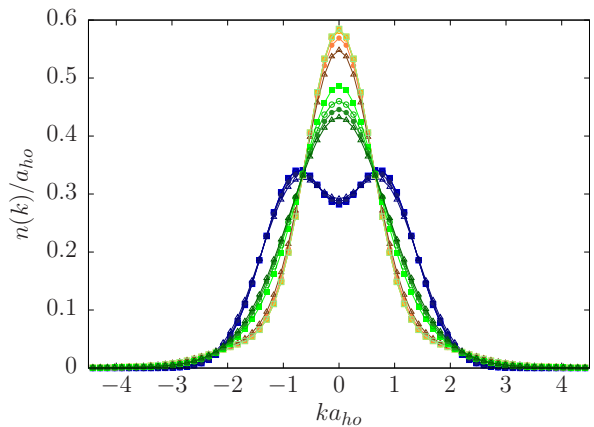}
    \caption{\label{figmom} Two-SU(2)-fermion momentum distribution
      $n(k)$ (green lines) compared with that for two Lieb-Liniger
      bosons (yellow lines) and two polarized fermions (blue lines),
      for $\xi_{\gamma}=7.06$, at different temperatures:
      $\tau$=2.5$\cdot10^{-3}$ (empty squares), 0.05 (full
      squares), 0.1 (empty circles), 0.15 (full circles), 0.2 (empty
      triangles).}
\end{center}
\end{figure}

\noindent
The momentum distribution is given by the Fourier transform of the
one-body density matrix:
\begin{equation}
  n(k)=\dfrac{1}{2\pi}\int {\rm d}x\int {\rm d}x'
  e^{-ik(x-x')} \rho(x,x'). 
\end{equation}
Analogously to the one-body density matrix, the momentum distribution
is a thermally weighted average of the momentum distribution of two
Lieb-Liniger bosons and the momentum distribution of
two spin-polarized fermions.  This is shown in Fig. \ref{figmom}.
At very low temperature, $k_BT/(\hbar\omega)$= 5$\cdot10^{-3}$, the
fermionic momentum distribution coincides with that for the
Lieb-Liniger gas, while as soon as the temperature increases there is
a hybridization between the momentum distribution of the Lieb-Liniger
gas and the spin-polarized fermionic one.
\subsubsection{The entanglement entropy.}

\begin{figure}
\begin{center}
 \includegraphics[width=1\linewidth]{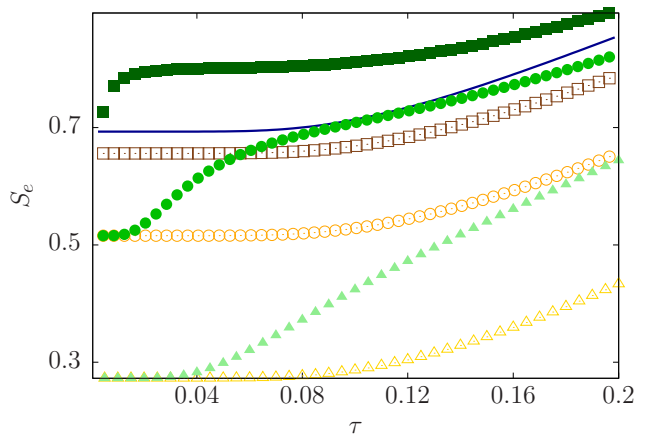}
 \caption{\label{fig-ent}Von Neumann entanglement entropy $S_e$ as a
   function of $\tau$ for different interaction strengths:
   $g/(\hbar\omega a_{ho})$=100 (squares), 10 (circles) and 3
   (triangles). The empty symbols correspond to Lieb-Liniger bosons,
   while the full symbols correspond to the case of SU(2) fermions.
   The continuous blue line marks the spin-polarized fermionic case.}
  \end{center}
\end{figure}

\noindent
One may wonder what the occurrence of this symmetry blending means from
the quantum information point of view.  To answer this question we
calculate the von Neumann entanglement entropy,
\begin{equation}
S_e=-{\rm tr}[\tilde\rho\ln(\tilde\rho)],
\end{equation}
where $\tilde\rho=\rho(x,x')a_{ho}$.  In Fig.  \ref{fig-ent} we plot
$S_e$ for two SU(2) fermions (full symbols) for different interaction
strengths: $g/(\hbar\omega a_{ho})$=100 (squares), 10 (circles) and 3
(triangles).  Each curve has to be compared with the entanglement
entropy for two Lieb-Liniger bosons (empty symbols) at the same
interaction strength, and with that for two spin-polarized fermions
(continuous line).

At small and intermediate interactions, the SU(2) curves are contained
between the Lieb-Liniger ones and the spin-polarized one.  But, at
very large interaction, approaching the Tonks limit, the SU(2) curve
overcomes the spin-polarized fermionic one.  Indeed, the finite
temperature Tonks limit corresponds to a maximal entanglement entropy:
the symmetric and the antisymmetric states becoming equiprobable, the
two fermions are maximally entangled.

For the spin-polarized case (blue continuous line), we recover at
$T=0$ the well-known limit $S_e=\ln(2)=0.693$ \cite{Santachiara2007},
while there is a sensible effect of the trap in the Tonks limit: in
the homogeneous gas it is expected $S_e=\ln(2)-0.30=0.393$
\cite{Santachiara2007}, while in the trapped system we find
$S_e=\ln(2)-0.037=0.656$ (empty squares). In addition, we find that
the sharp increase of the entanglement entropy at low temperature and
the minimum of the contact occur simultaneously around a temperature
$T_0$.
\section{Conclusions}
\label{concl}
In this paper we have studied the Tan's contact for $N$ harmonically
trapped 1D SU(2) fermions characterized by repulsive contact
interactions.  By means of a LDA calculation we have verified that the
Tan's contact exhibits a minimum at very low temperature as expected
in the homogeneous system \cite{Patu2016}.  With the aim to improve
the understanding of the contact minimum, we have calculated the
two-fermion contact as well.  At $T=0$ the fermionic contact coincides
with that of two indistinguishable bosons since the ground state is
totally symmetric, while at high temperature the contact is equal to
half of the bosonic one since the symmetric and antisymmetric
components have almost the same statistical weight. The minimum, that
is a signature of this thermally driven symmetry blending, occurs at
an energy scale determined by the energy difference between the ground
state and the first excited state. We find that this difference is
proportional to the ground-state contact in the large interaction
limit.  Moreover, we have shown that the symmetry blending, that can
be observed in other observables such as the momentum distribution, in
the strongly interacting limit, corresponds to a maximal entanglement.

\begin{acknowledgments}
P.V. acknowledges O. P\^a\c{t}u for very useful exchanges, and
F. Chevy, C. Salomon, F. Werner, J. Decamp, M. Albert for useful
discussions. The authors also acknowledge A. Minguzzi for her
suggestions during the early stages of this work.  This research has
been carried out in the International Associated Laboratory (LIA)
LICOQ.  P. C. acknowledges partial support from CONICET and Universidad de
Buenos Aires through grants PIP 11220150100442CO and UBACyT 20020150100157,
respectively.

\end{acknowledgments}
\appendix
\section*{Appendix: The one-body density matrices of two particles}
In this section we give some details on the calculation of the
one-body density matrices for the antisymmetric and symmetric cases
for two SU(2) fermions.

The antisymmetric contribution corresponds to purely noninteracting
fermions, and as such the expression for the  one-body
density matrix is well-known for arbitrary $N$. For two fermions it
can be written as functions of the two occupied single-particle
states $i$ and $j$ as
\begin{equation} 
\rho_a^{i,j}(x,x') = \frac{1}{2}\left(\varphi_i(x)\varphi_i(x') + \varphi_j(x)\varphi_j(x')\right).
\end{equation}
Given that the total energy for this state is $E_{i,j}^a$, the sum
over $i,j$ entering Eq. (\ref{eq:rdmT}) can be exactly performed in
the harmonic confinement case thanks to Mehler's formula
\cite{Foata1978,Watson1933} which states that
\begin{multline}
  \mathcal{K}(x,x',u) \equiv  \sum_{n=0}^{\infty} \varphi_n(x)\varphi_n(x') u^n = \\
  \dfrac{1}{\sqrt{\pi (1 - u^2)}}\exp\left\{-
    \frac{1}{4}\left[\dfrac{1 - u}{1 + u} (x + y)^2 + \dfrac{1 + u}{1
        - u} (x - y)^2\right]\right\}
\end{multline}
where
$\varphi_n(x)=H_n(x/a_{ho})/\sqrt{a_{ho}2^nn!\sqrt{\pi}}\,e^{-m\omega
  x^2/2\hbar}$ is the normalized 1D harmonic oscillator eigenfunction
and $|u|<1$. $H_n(x)$ is the Hermite polynomial of order
$n$. Therefore, summing up the different terms on $\rho_a^{i,j}$ in
(\ref{eq:rdmT}) one finds
\begin{equation}
  \sum_{i,j} e^{-\beta E_{i,j}^a}\rho_a^{i,j}(x,x') = \mathcal{K}(x, x', u_\beta)\,
  \frac{u_\beta}{1 - u_\beta} - \mathcal{K}(x, x', u_\beta^2)\,u_\beta
\label{eq:rdm1a}
\end{equation}
where $u_\beta=e^{-\beta\hbar\omega}$. This expression allows to
obtain an analytical formula for the corresponding density at finite
temperature
\begin{widetext}
\begin{align}
  \rho_a(x) =& \frac{e^{\beta \hbar\omega}e^{-x^2 \left(\tanh \beta  \omega  \hbar/2 +\tanh \beta  \omega  \hbar\right)}}{2 \sqrt{\pi }
              \sqrt{e^{2 \beta  \omega  \hbar }+1}}   \left\{ e^{\beta \hbar\omega}\sqrt{1-e^{-4\beta\hbar\omega}}\left(\cosh(x^2\tanh\beta\hbar\omega)+ \sinh(x^2\tanh\beta\hbar\omega)\right) \right. \nonumber \\ &\left.- (e^{\beta\hbar\omega}-1)\sqrt{1-e^{-\beta\hbar\omega}}e^{x^2\tanh\beta\hbar\omega}\right\}.
\end{align}
\end{widetext}

The symmetric contribution to the one-body density matrix is more
involved as it explicitly depends on $g$ through
$\nu(i)$. Nevertheless, the summation over the center-of-mass degrees
of freedom can be exactly performed as in the antisymmetric case
leading to an expression with a single sum left
\begin{multline}
  \sum_{i,j} e^{-\beta E_{i,j}^s}\rho_s^{i,j}(x,x') = \sum
  e^{-\beta\hbar\omega(1+\nu(i))} \\ \times \int {\rm d}y\,
  \mathcal{K}\left( \frac{x+y}{2a_X},
    \frac{x'+y}{2a_X},u_\beta\right)\phi_i(x-y)\phi_i(x'-y)
  \label{eq:rdm1s}
\end{multline}
where
\begin{equation}
  \phi_i(x)=\frac{1}{\sqrt{\mathcal{N}_{\nu(i)} a_x}}\mathop
  U\left(-\frac{\nu(i)}{2},\frac{1}{2},\frac{x^2}{2 a_{ho}^2}\right)e^{-m\omega x^2/4\hbar}
\end{equation}
is the normalized eigenfunction of the relative motion with energy
$\hbar\omega(1/2+\nu(i))$ (c.f. Eq.\ (\ref{eq:Erel})) and normalizing
constant \cite{Rizzi2018}
\begin{multline}
  \mathcal{N}_\nu = 2^{-\nu}\Gamma(\nu+1)\sqrt{\pi}\\\left(1+\frac{\sin\pi\nu}{2\pi}\left(\psi(\nu/2+1)-\psi(\nu/2+1/2)\right)\right).
  \label{eq:Nca}
\end{multline}
The functions $\mathop U$ and $\psi$ are the confluent Hypergeometric
function and the digamma function, respectively.  Combination of the
symmetric and antisymmetric expressions above together with their
canonical partition functions allows to efficiently calculate the
one-body density matrix for the two fermions, their momentum
distribution and von Neumann entropy.

\end{document}